\documentclass[12pt]{iopart}

\pdfoutput=1
\usepackage[breaklinks=true,colorlinks=true,linkcolor=blue,urlcolor=blue,citecolor=blue]{hyperref}
\usepackage{latexsym}
\usepackage{graphics}
\usepackage{graphicx}
\usepackage{epsfig}
\usepackage{mathcomp}
\usepackage{amsopn}
\usepackage{amsfonts}
\usepackage{amstext}
\usepackage{amssymb}
\usepackage{amstext}
\usepackage{subfigure}
\usepackage{braket}
\usepackage{iopams}  
\usepackage{wasysym}

\usepackage{epstopdf}
\usepackage{braket}
\begin{document}

\title{{\em Ab initio} calculation of Hubbard parameters for Rydberg-dressed atoms in a one-dimensional optical lattice}

\author{Yashwant Chougale and Rejish Nath}

\address{Indian Institute of Science Education and Research, Pune 411 008, India}
\vspace{10pt}

\begin{abstract}
We obtain {\em ab initio} the Hubbard parameters for Rydberg-dressed atoms in a one-dimensional sinusoidal optical lattice in the basis of maximally localized Wannier states. Finite range, soft-core inter-atomic interactions become the trait of Rydberg admixed atoms, which can be extended over many neighbouring lattice sites. On contrary to dipolar gases, where the interactions follow an inverse cubic law, the key feature of Rydberg-dressed interactions being the possibility of making neighbouring couplings to the same magnitude as that of the onsite ones. The maximally localized Wannier functions are typically calculated via spread minimization procedure [Phys. Rev. B {\bf 56}, 12847 (1997)] and always found to be real functions apart from a trivial global phase when considering an isolated set of Bloch bands. For an isolated single Bloch band, the above procedure reduces to a simple quasi-momentum dependent unitary phase transformation. Here, instead of minimizing the spread, we employ a diagonal phase transformation which eliminates the imaginary part of the Wannier functions. The resulting Wannier states are found to be maximally localized and in exact agreement with those obtained via spread minimization procedure. Using that we calculate the Hubbard couplings from the Rydberg-admixed interactions, including dominant density assisted tunnelling coefficients. In the end we provide realistic lattice parameters for the state of the art experimental Rydberg dressed Rubidium setup.
\end{abstract}

%
%
\submitto{\JPB}
%
\maketitle
%
%
\tableofcontents
\section{Introduction}
A new wave of study has emerged in cold Rydberg atom-community along the lines of so-called {\em Rydberg-dressed atoms} \cite{sant00, hon10, john10, pup10, hen10, tom14, bal14}. They are mostly the ground state atoms with a very tiny fraction of Rydberg excited ones, represented by the state vectors of the form $\sim |g\rangle+\alpha|r\rangle$ with $|\alpha|\ll1$, where $|g\rangle$ and $|r\rangle$ are respectively the ground and Rydberg states. Rydberg excited atoms are known to exhibit prodigious inter-atomic interactions \cite{saf10} that can suppress further excitations within a finite volume, called the Rydberg blockade \cite{luk01, urb09, gae09}. Though the excitations are short lived, the huge interactions made it feasible to study certain interesting many-body effects, within the frozen gas limit \cite{poh10, wei08, low09, wei10, olm09, ji11, igo11}. Later, Rydberg admixing turned up as a remedy to overcome the lifetime constraint of Rydberg excited atoms, especially augmented its effective lifetime by a factor of $1/|\alpha|^2$. Interatomic interactions with a softcore potential barrier \cite{hen10} become the trait of Rydberg-dressed atoms and is verified in two recent experiments. In one of them, the effective interactions in the strong dressing limit are demonstrated with two individual atoms trapped in optical tweezers \cite{jau15} where as the second experiment involves a two-dimensional (2D) lattice setup \cite{zei16}. In a short span of time, it has led to various exciting studies: in quantum many-body physics \cite{hen10, hsu12, cin10, li15, gei15, mau11,ric15,dau12} including frustrated quantum magnets \cite{ale14, ale15}, quantum computing \cite{kea13,kea15,mob13} and spin squeezing for metrology \cite{gil14, bou02}. As lately proposed, an electro-magnetically induced transparency based Rydberg dressing scheme utilizing metastable states of Alkaline earth atoms may help to achieve enhanced atomic interactions even with longer coherence times \cite{gau15}.

The quantum simulation of many-body physics predominantly relies on ultra cold atoms loaded in optical lattices \cite{lew12,blo08} described by local Hubbard models \cite{jak98}. The Bose-Hubbard model (BHM) with contact interactions and nearest neighbour hoppings forms the paradigmatic example of a system exhibiting quantum phase transition, dating back to the prediction of  superfluid-mott insulator transition by Fisher {\em et. al} \cite{fis89} and for the first time it is observed in a cold atom-lattice setup \cite{gre02}.  Due to the tremendous progress in experimental techniques, in particular, the ability to address a single atom in a lattice site  \cite{ger08, bak09, she10} as well as to describe the experimental results quantitatively accurate it is required to calculate the Hubbard parameters on a basis of highly localized single particle states. An example being the  basis of maximally localized Wannier states, can be obtained by unitary mixing of the Bloch states \cite{koh59,bro07}. In a seminal work by  Marzari and Vanderbilt \cite{mar97}, they showed how to calculate the maximally localized Wannier functions (MLWFs) by minimizing the spread of generalized Wannier states. Their approach has been employed successfully in studying complex solid state materials \cite{mar12} and recently been used for cold atoms in optical lattices \cite{wal13}. When lattice depths are sufficiently large, tight binding together with harmonic approximation is in good agreement with MLWF calculations for lowest isolated bands.

The well-acclaimed BHM is then generalized to systems with long-range interactions (e.g. magnetic atoms, polar molecules or atoms with laser-induced interactions) termed as an extended BHM (EBHM) \cite{bar12, tre11} in which the interactions between neighbouring sites are included. EBHM is predicted to be abundant with exotic quantum phases such as stripes, checkboard phases, supersolids, Haldane insulators etc \cite{sca95, sca05, yi07, pol10, cap10, dall06}. Lately, it has been demonstrated and experimentally probed using magnetic Erbium atoms trapped in a 3D optical lattice \cite{bai15}. Here, we derive the EBHM in a 1D lattice of Rydberg admixed atoms from first principles and as we show it has certain advantages over dipolar systems. Among them, the most striking property is the possibility of engineering the strengths of off-site couplings to the same magnitude as that of onsite ones, hence position independent. This may have far-reaching consequences in the context of quantum many-body physics \cite{matt13} especially in frustrated magnetism \cite{ale14, ale15} to impose local constraints or conservation laws. 

An additional  correlated feature of interacting lattice gases is the density assisted inter-site or inter-band tunneling, in which the former has been observed in atoms with both contact interactions \cite{jur14} and long-range dipolar interactions \cite{bai15}. Termed as bond-charge interaction in the context of fermions, the density assisted tunneling (DAT) strongly influences the MOT insulator-superfluid transition points in bosonic and Bose-Fermi mixtures \cite{dut15} and also lead to novel quantum phases for polar molecules in optical lattices \cite{sow12}.

Here, we calculate {\em ab initio} the Hubbard parameters on the basis of MLWFs for an extended Hubbard model implemented using Rydberg-dressed atoms in a one-dimensional sinusoidal optical lattice. Motivated by the hypothesis that MLWFs are real functions apart from a trivial global phase, we implement a diagonal $U(1)$ phase transformation for the Bloch states which eliminate the imaginary part of the Wannier states. The resulting Wannier states are found to be maximally localized and  in exact agreement with those obtained via spread minimization procedure. We estimate the Hubbard parameters including the DAT coefficients and discuss their properties within the two lowest Bloch bands and up to second nearest  neighbour couplings for Rydberg admixed interaction potential. 

The paper is structured as follows: in section \ref{setup} we discuss the Rydberg-lattice setup, the governing many-body Hamiltonian and derive {\em ab initio} the extended Hubbard model in the maximally localized Wannier basis. In section \ref{mlwf} we discuss how MLWFs are calculated using a unitary diagonal phase transformation in the isolated Bloch bands for a sinusoidal 1D optical lattice potential. The results for the Hubbard parameters are discussed in section \ref{results} as well as we provide realistic numbers for the state of the art experimental Rubidium lattice setup \ref{rb}. Finally, we conclude in section \ref{con}.

\section{Rydberg dressed atoms in an one-dimensional optical lattice}
\label{setup}
\subsection{The atom-lattice setup and Rydberg-Rydberg interactions}
\begin{figure}[hbt]
\centering
\includegraphics[width= .75\columnwidth]{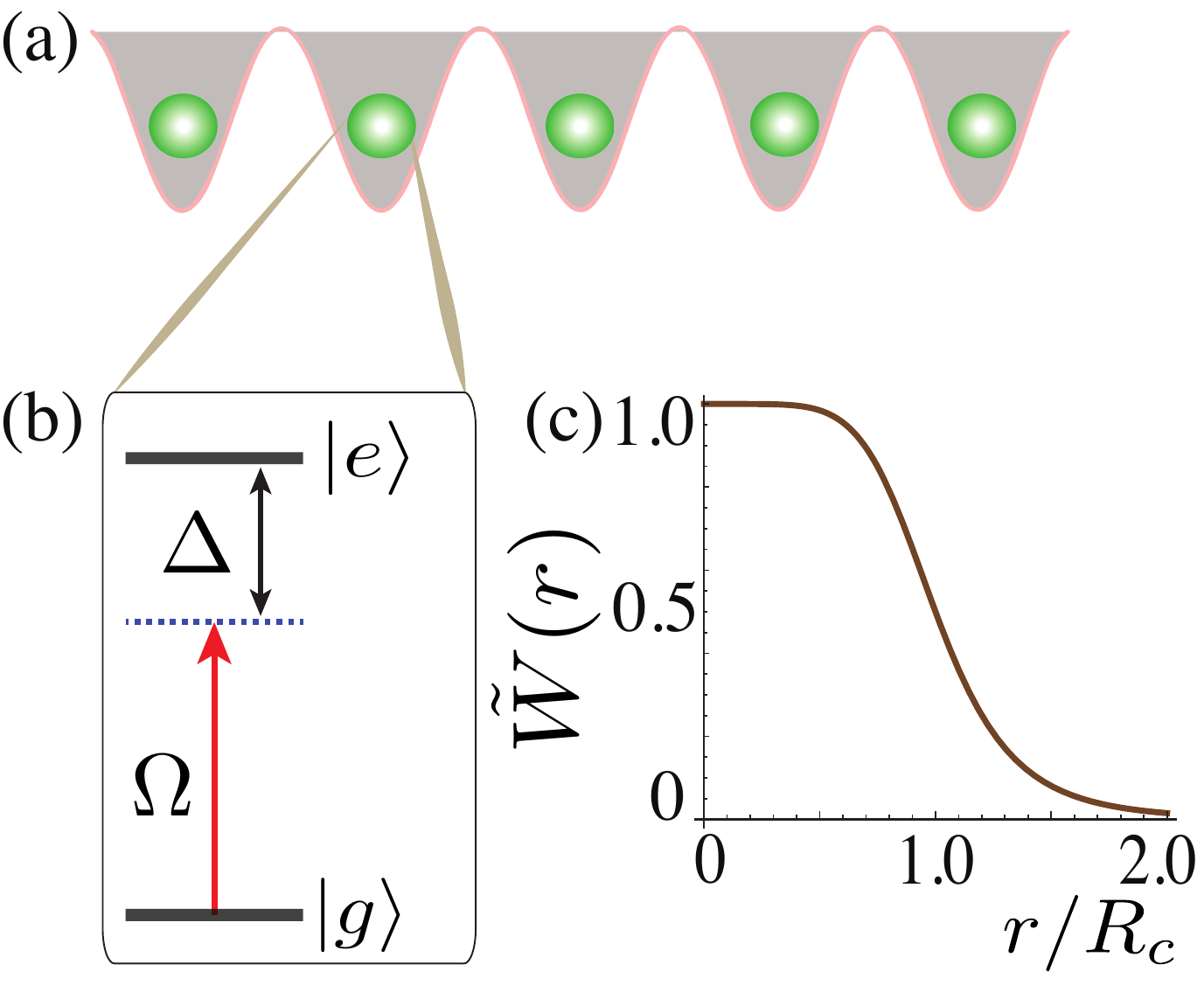}
\caption{\small{(a) The one-dimensional lattice setup with Rydberg dressed atoms. (b) The two-level scheme which consists of a ground state $|g\rangle$ being weakly coupled to a Rydberg state $|e\rangle$ by a laser field with  detuning $\Delta$ and  Rabi frequency $\Omega$. (c) The emerging inter-atomic interactions  between the dressed ground state atoms: $\tilde W(r)=W(r)/W_{eff}$ [see equation (\ref{dpotn})] as a function of interparticle separation $r$.}}
\label{fig:setup} 
\end{figure}

We consider bosonic atoms loaded in a one-dimensional optical lattice of potential $V_{OL}(x)=V_0\sin^2(k_0x)$, where $V_0$ is the lattice depth and wavenumber $k_0=\pi/d$ with $d$ being the lattice spacing.  The ground state $|g\rangle$ of the atoms is weakly admixed to an excited Rydberg  state $|e\rangle$ using an optical field with a large detuning $\Delta$ compared to Rabi frequency $\Omega$ such that $\Omega/|\Delta|\ll 1$, see Figure \ref{fig:setup}.  We assume that both ground and Rydberg state atoms are trapped in a magic wavelength optical lattice (MWOL) such that atoms occupying both states experience identical lattice potentials \cite{saf03, saf05, muk11, gol15}. Large wavelength infrared MWOLs due to effective landscape polarizability, suitable for high lying alkali atom Rydberg states are also proposed \cite{top13}.  If the Rydberg state trapping is neglected an additional time constraint arises due to the expansion of the atomic wave packet (see \ref{val} for details) when atoms occupy the Rydberg state. Two atoms with an interatomic separation $r$ such that away from F\"orster regime, occupying the state $|e\rangle$ experience an interaction potential of the form $C_6/r^6$, where the van der Waals (vdW) dispersion coefficient $C_6(\propto n^{11})$ can be either spatially isotropic or anisotropic \cite{ale14,mau11} depending on different angular momentum quantum numbers $\{l, j, m_j\}$ of the Rydberg state $|e\rangle\equiv |n, l, j, m_j\rangle$, where $n$ is the principal quantum number. The above vdW interactions between the Rydberg excited atoms lead to a tunable soft-core potential for the atoms in the admixed ground state (see Figure \ref{fig:setup}c) and is 
\begin{equation}
W(r_{ij})=\left(\frac{\Omega}{2|\Delta|}\right)^4\frac{C_6}{(r_{ij}^6+R_c^6)}.
\label{dpot}
\end{equation}
 The parameter, $R_c=[C_6/2\hbar|\Delta|]^{1/6}$ being the Condon radius determines the range of the interactions and $r_{ij}$ is the spatial separation between the $i^{th}$ and $j^{th}$ atoms. The Rydberg blockade effect is apparent at distances $r_{ij}\leq R_c$, where the interaction potential is saturated to a constant value provided by the light shift of the blockaded atoms \cite{hen10,jau15}. For large values of $r_{ij}>R_c$, we retrieve the vdW type interactions, $W(r_{ij})\approx \tilde C_6/r_{ij}^6$ with a reduced vdW coefficient $\tilde C_6=(\Omega/2|\Delta|)^4C_6$. Notably, the range of the interactions $\sim R_c$ can be extended over several lattice sites by regulating the detuning $\Delta$ or the interaction coefficient $C_6$, fabricating the scenario identical to that of long-range interacting systems such as magnetic atoms or polar molecules \cite{tre11, bai15, sow12}. 
  
\subsection{Many body Hamiltonian and the extended Hubbard model}
Our starting point is the second quantization many-body Hamiltonian for a gas of interacting bosons in an external potential $V_{OL}({\bf r})$:
\begin{eqnarray}
\hat H=\int d{\bf r} \ \hat\Psi^{\dagger}({\bf r})\hat h_0\hat\Psi({\bf r})+\frac{1}{2}\int d{\bf r}\int d{\bf r}^{\prime} \ \hat\Psi^{\dagger}({\bf r})\hat\Psi^{\dagger}({\bf r}^{\prime})U_I({\bf r}-{\bf r}^{\prime})\hat\Psi({\bf r})\hat\Psi({\bf r}^{\prime}),
\label{h2nd}
\end{eqnarray}
where the field operator $\hat\Psi({\bf r})$  [$\hat\Psi^{\dagger}({\bf r})$] annihilates [creates] a boson at the position ${\bf r}$. The term
$\hat h_0=-\hbar^2\Delta^2/2m+V_{OL}({\bf r})$ corresponds to the single particle Hamiltonian. The two-particle interactions are, $U_I({\bf r}-{\bf r}^{\prime})=U({\bf r}-{\bf r}^{\prime})+W({\bf r}-{\bf r}^{\prime})$ with the first term being the typical contact interactions: $U({\bf r}-{\bf r}^{\prime})=g\delta({\bf r}-{\bf r}^{\prime})$ characterized by the parameter, $g=4\pi\hbar^2a/m$ where $a$ is the $s$-wave scattering length, and the second term is the interaction potential among the Rydberg-admixed atoms. Next, we expand the field operators in the basis of maximally localized single particle Wannier functions $w_n({\bf r}-{\bf R}_j)$ (see section \ref{mlwf}) centered around each lattice site at ${\bf R}_j=j{\bf R}$ and the lattice vector ${\bf R}$ is  such that $V_{OL}({\bf r}+{\bf R})=V_{OL}({\bf r})$. Hence, the field operator
\begin{equation}
\hat\Psi({\bf r})=\sum_{j}\sum_n w_n({\bf r}-{\bf R}_j)\hat a_{jn},
\label{fop}
\end{equation}
where $\hat a_{jn}$ ($\hat a_{jn}^{\dagger}$) annihilates (creates) a boson in the mode represented by the Wannier function $w_n({\bf r}-{\bf R}_j)$ at $j^{th}$ lattice site. They satisfy the commutation relation $[\hat a_{in}, \hat a^{\dagger}_{jm}]=\delta_{ij}\delta_{nm}$. Inserting the above expression for field operator in Equation \ref{h2nd}, we arrive at a general extended  Hubbard model:
\begin{eqnarray}
\hat H=-\sum_{i,j}\sum_{n,m}J_{ij}^{nm}\hat a_{in}^{\dagger}\hat a_{jm}+\frac{1}{2}\sum_{i,j,i',j'}  \sum_{n,m,n',m'} (U_I)_{iji'j'}^{nmn'm'}\hat a_{in}^{\dagger}\hat a_{jm}^{\dagger}\hat a_{j'm'}\hat a_{i'n'}
\label{ebhm}
\end{eqnarray}
where the hopping and the interaction matrix elements are respectively
\begin{equation}
J_{ij}^{nm}=-\int d{\bf r} \ w_n^*({\bf r}-{\bf R}_i)\left[-\hbar^2\Delta^2/2m+V_{OL}({\bf r})\right]  w_m({\bf r}-{\bf R}_{j}) 
\label{hop}
\end{equation}
\begin{equation}
(U_I)_{iji'j'}^{nmn'm'}=\iint d{\bf r}d{\bf r'} \ w_n^*({\bf r}-{\bf R}_i) w_m^*({\bf r'}-{\bf R}_j)U_I({\bf r}-{\bf r'})w_{m'}({\bf r'}-{\bf R}_{j'})w_{n'}({\bf r}-{\bf R}_{i'}).
\label{int}
\end{equation}
The subscripts and superscripts in $J$ and $U_I$ represent the lattice site and band indices respectively. In cold atoms, it can be the case that interaction and kinetic energies are smaller than the lattice depth $V_0$, then the summation over the band indices is truncated beyond few lowest Bloch bands. Below we calculate the microscopic parameters for the EBHM (equation \ref{ebhm}) up to next nearest neighbour couplings within the two lowest Bloch bands for the 1D lattice setup.


\subsection{Validity of our model}
\label{val}
{\em Rydberg state trapping}:- As mentioned before, our calculations we provide below are based on the assumption that both ground and Rydberg state atoms experience the same lattice potential which requires MWOL. If it is not the case motional heating can occur during ground-Rydberg state transitions, which may lead to severe decoherence \cite{li13}. The Rydberg-Rydberg interactions can also induce mechanical effects at short distances and that can be safely neglected for Rydberg-dressed atoms \cite{tom14}. 
In this section, we discuss what criteria in which the effective interactions calculated using Equation (\ref{int}) are valid when no-trapping is provided for Rydberg atoms. It depends crucially on the timescale at which the free expansion of atomic wave packet results in a considerable change of the Wannier state. It has to be shorter compared to the lifetime of Rydberg atoms and the timescales set by effective interaction strengths in the lattice. 

 Assuming the harmonic oscillator (HO) states for the two lowest Wannier states (valid in the tight binding case), we estimate the duration at which the wave-packets remain intact after excited to a trap-free Rydberg state. The time evolution of the standard deviation for the widths in the $n^{th}$ state of the HO is given by
\begin{equation}
\Delta X_n(t)/\Delta X_n(0)=\sqrt{1+\frac{\hbar^2 t^2}{m^2l_0^4}},
\end{equation}
where $l_0=\sqrt{\hbar/m\omega}$. The frequency $\omega$ of the HO in terms of the lattice depth $V_0$ and lattice spacing $d$ is $\omega=(\pi/d)\sqrt{\frac{2V_0}{m}}$. Note that we are interested in a duration of time $\tau$ such that the wave packet hardly undergoes any expansion, i.e. when the criteria:
\begin{equation}
\frac{\hbar^2 \tau^2}{2m^2l_0^4}\ll 1 \ \ \ \ \ \ \ {\rm or} \ \ \ \ \ \ \ \ \ \  \frac{\tau^2 V_0\pi^2}{md^2}\ll 1
\label{cri}
\end{equation}
is satisfied. For a given $V_0$ equation (\ref{cri}) can be satisfied by sufficiently large values of $d$ and it is also highly desirable for us since we want to access the van der Waals regime for the interactions between the Rydberg atoms. The large lattice spacings of the order of micrometers can be accessed by adjusting the angle $\theta$ between the co-propagating lattice beams \cite{nel07}, given as $d=\lambda/[2\sin(\theta/2)]$. In section \ref{rb}, we discuss the criteria in equation \ref{cri} for the case of a Rubidium lattice setup.

{\em Rydberg-Rybderg interactions} :- For two atoms occupying the same Rydberg state $|\alpha\rangle$ and far away from F\"orster resonance the second order level shifts due to the dipole-dipole interactions to the pair state $|\alpha\alpha\rangle$ provide us the vdW interactions between the atoms. The F\"orster resonance is characterized by the F\"orster defect: $\delta_F=E_{\alpha\alpha}-E_{\beta\gamma}$, where the atomic states $|\beta\rangle$ and $|\gamma\rangle$ are dipole coupled to the state $|\alpha\rangle$. This introduces a cross-over between F\"orster regime at short distances and vdW regime at large separations between the atoms. The transition point $R_F$ can be calculated as
\begin{equation}
R_F^3=\sum_{\beta, \gamma}\left|\frac{\langle \beta \gamma|V_d(R)|\alpha\alpha\rangle}{\delta_F}\right|,
\label{rc}
\end{equation}
where $V_d(R)$ is the dipole-dipole interaction and hence, the lattice spacing has to be greater than $R_F$ for our calculations to be valid. We estimate $R_F$ for Rubidium $nS_{1/2}$ states in section \ref{rb}.

\section{Maximally localized Wannier functions}
\label{mlwf}
The generalized Wannier functions are calculated by the unitary mixing of $N$ degenerate or closely spaced Bloch eigenstates and in 1D it is written as,
\begin{equation}
 w_n(x-X_j)=\frac{d}{2\pi}\int_{BZ}dq \ e^{-iqX_j}\sum_{m=1}^NU_{nm}(q)\psi_q^m(x),
\end{equation}
where $\psi_q^m(x)$ is the Bloch function for the $m^{th}$ Bloch band for a quasi-momentum $q$. The operator $U_{nm}(q) \in U(N)$ is a unitary matrix satisfying the periodic boundary condition in the momentum space: $U_{nm}(q+2q_B)=U_{nm}(q)$ with $q_B=\pi/d$ and $N$ corresponds to the number of lattice sites in the unit cell. $BZ$ indicates that the integration is carried over the first Brillouin zone. As proposed in a seminal paper, the MLWFs can be obtained from generalized Wannier functions by obtaining $U_{nm}(q)$ which minimizes the spread functional \cite{bro07}:
\begin{equation}
\Omega_w=\sum_{n=1}^N\left[\langle 0n|x^2|0n\rangle-\langle 0n|x|0n\rangle^2\right],
\end{equation}
where $\langle x|jn\rangle=w_n(x-X_j)$. MLWFs are shown to be exponentially localized \cite{koh59,he01} and they provide the best optimal basis for estimating Hubbard parameters. For an isolated single Bloch band ($N=1$), the unitary transformation is an abelian $U(1)$ gauge transformation and, is nothing but an update for the phase of Bloch functions i.e. $\psi_q^m \to e^{i\phi_q^m}\psi_q^m$. It is also the case for the setup we considered here.  Thus, once the Bloch spectrum (see \ref{bbs}) is found, the problem for calculating MLWFs reduces to finding the phases $\phi_q^m$ which minimize the spread $\Omega$. It has been shown that the results obtained by minimizing the Wannier spread for isolated Bloch bands are same as the exponentially localized wannier functions discussed by Kohn \cite{koh59}. Writing the Wannier spread as a sum of gauge independent and dependent terms: $\Omega_w=\Omega_I+ \tilde \Omega_D$, with subsequent division of gauge dependent one to diagonal and off-diagonal terms: $\tilde\Omega_D=\Omega_d+ \Omega_{od}$. For an isolated band $\Omega_{od}$ vanishes and the question reduces to finding the unitary transformation which eliminates $\Omega_d$ \cite{bro07}. Note that, there also exists other localization procedures for Wannier states e.g. maximizing the sum of Coulomb self-energies \cite{edm63}.

Here we do not employ the minimization of Wannier spread, rather make use of the conjecture that MLWFs are real functions up to a global phase factor. In our case ($N=1$) the generalized Wannier functions are
\begin{equation}
 w_n(x-X_j)=\frac{d}{2\pi}\int_{BZ}dq \ e^{-iqX_j} \  e^{i\phi_q^n} \ \psi_q^n(x).
 \label{w2}
\end{equation}
The periodic Bloch functions $\psi_q^n(x)$ are calculated as prescribed in \ref{bbs}. The phases $\phi_q^n$ for the unitary transformation is obtained by minimizing the absolute value of the imaginary part of the Wannier function in equation \ref{w2} and it leads to 
\begin{equation}
\phi_q^n=\arctan\left(\frac{\Re[u_q^n(x_0)]}{\Im[u_q^n(x_0)]}\right)-qx_0.
\end{equation}
In our numerics, for symmetric $u_q^n(x)$ we took $x_0=0$ and for anti-symmetric $u_q^n(x)$ we took the point at which the imaginary part of $u_q^n(x)$ is maximum. The resulting Wannier functions match perfectly with those obtained via spread minimization procedure \cite{wal12}. As an example, MLWFs for the two lowest bands ($n=1, 2$) in a 1D sinusoidal lattice are shown in figure \ref{fig:uj}a for $V_0=10E_R$ and their exponential localization is conspicuous in the log scale plot of the vertical axis. 


\section {Hubbard Parameters}
\label{results}
\subsection{Hopping and short-range interaction parameters}
\label{si}
\begin{figure}[hbt]
\centering
\includegraphics[width= .85\columnwidth]{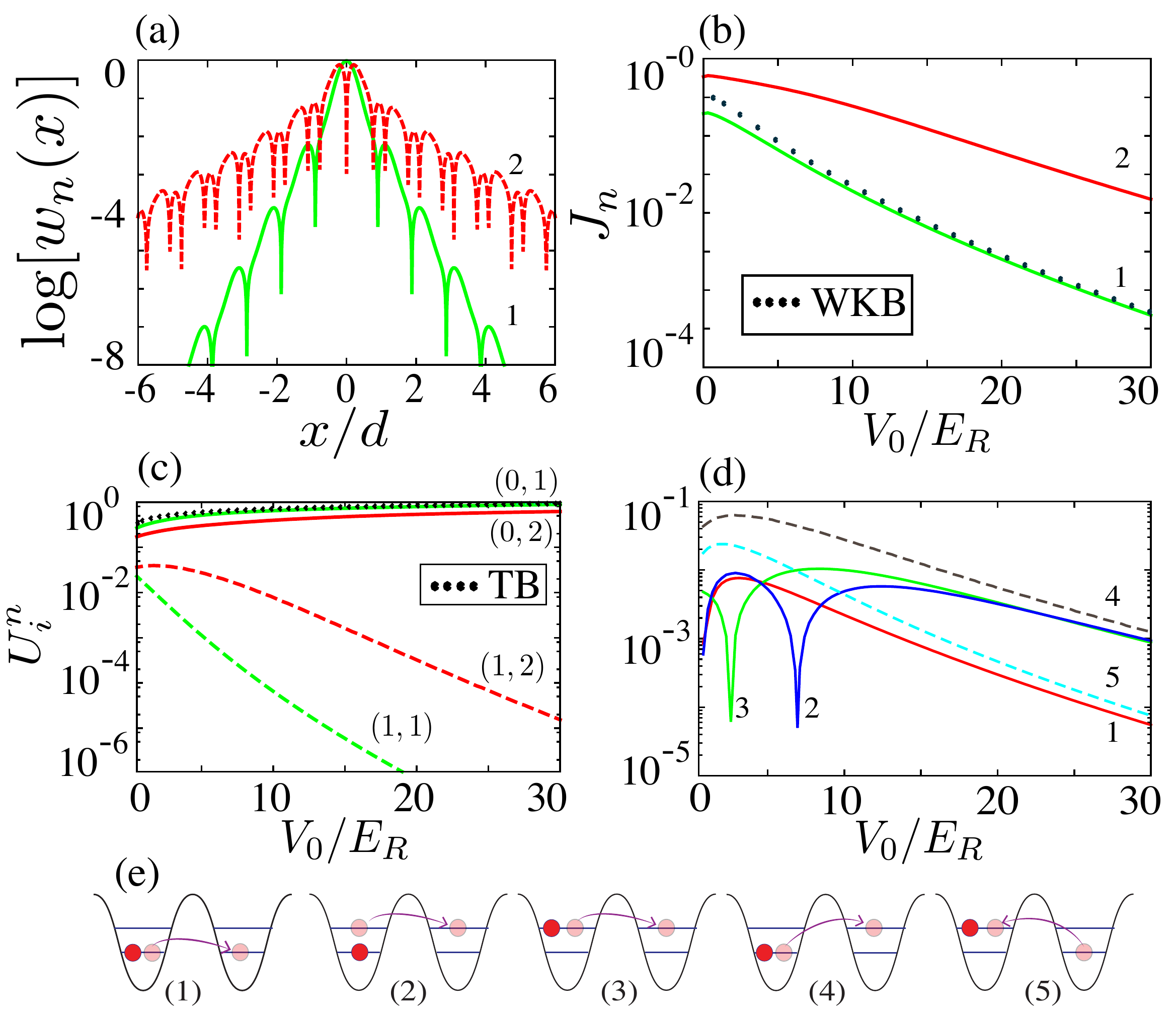}
\caption{\small{(a) The maximally localized Wannier functions $w_n(x)$ for the two lowest Bloch bands in a 1D sinusoidal optical lattice with $V_0=10E_R$. The exponential decay of MLWFs is clearly visible with a log-scale along the $y$-axis. (b) The nearest neighbour hopping parameter $J_n=J_{12}^{nn}$ (see equation. \ref{hop}) for the two lowest bands as a function of lattice depth $V_0$. As expected, the hopping get suppressed as $V_0$ becomes larger and larger. Note that Higher the band index larger the hopping matrix element for a given $V_0$ as evident from the Wannier functions that their spatial extension increases with $n$. For $n=1$, the results from the WKB approximation (dotted line) is in good agreement, but for higher bands (not shown) they differ considerably. The band indices for figures (a) and (b) are shown in the corresponding plots. (c) The on-site (solid lines and $i=0$) and nearest neighbour (dashed lines and $i=1$) short range interaction parameters  ($U_i^n=U_{iiii}^{nnnn}$) for the two lowest bands ($n=1, 2$). Onsite interaction is higher for the lowest band and the nearest neighbour interactions are opposite to it. The results from the tight-binding approximation \cite{wal12} for the lowest band are shown on the dotted line. (d) shows the scaled (units of $\tilde g$) DAT amplitudes as a function of $V_0$, which include DAT within bands to the nearest neighbour site (solid lines) and those between bands to the nearest neighbour site (dashed lines). The different DAT processes are shown in (e) and the corresponding amplitudes are indicated by the same numbers in (d). }}
\label{fig:uj} 
\end{figure}

We briefly summarize the behaviour of Hopping ($J$) and contact interaction ($U$) matrix elements as a function of lattice depth $V_0$ calculated using MLWFs (see figure \ref{fig:uj}). The spatial spread of Wannier functions $w_n(x)$ gets  larger and larger as we go higher in bands ($n$) and for a given $n$ it decreases with increasing $V_0$. The localization properties of $w_n(x)$ are critical in determining the behaviour of $J$ and $U$. For instance, the nearest neighbour hopping ($J_n^0=J_{12}^{nn}$) is larger in the second band compared to that in the first one. Also for both bands, it decreases exponentially with increasing $V_0$ [figure \ref{fig:uj}(b)]. The next nearest neighbour hopping is two orders of magnitude less than $J_n$ and is safely disregarded in the (E)BHM.  Note that there is no inter-band hopping  due to the orthogonality property of Wannier functions. The results from WKB approximation \cite{wal12} for the lowest ($J_1^0)$ is shown in dotted line in \ref{fig:uj}(b) and is in good agreement with that of exact results. 

Since $w_n(x)$ gets more and more localized with larger values of $V_0$, the onsite interactions ($U_0^n=U_{0000}^{nnnn}$) for any $n$ increases as a function of $V_0$ [figure \ref{fig:uj}(c)]. As expected, the behaviour for the nearest neighbour interactions ($U_1^n=U_{0101}^{nnnn}$) is just opposite and also few orders of magnitude lower than that of the onsite ones, which may hardly affect the many body phase diagram of BHM. On the other hand, the different DAT matrix elements shown in figure \ref{fig:uj}(d), though their magnitudes are relatively small (has to be compared with hopping amplitudes $J_n$) some of them may significantly modify the dynamics in BHM \cite{jur14}, even in sufficiently strong optical lattices. In particular, if restricted to the lowest band the dominant off-site contribution in BHM comes from a DAT term [first processes in figure \ref{fig:uj}(e)] of the form $\sim \sum_i\hat a^{\dagger}_i(\hat n_i+\hat n_{i+1})\hat a_{i+1}$ and is explicitly occupation dependent.  The signatures of this term in a deep MOT state are probed in a recent experiment using tilted optical lattices \cite{jur14}.  In figure \ref{fig:uj}(d) only the dominant DAT coefficients are shown and corresponding processes are depicted in figure \ref{fig:uj}(e). Note that onsite inter-band tunnelings are prohibited by the orthogonality property of MLWFs. With the inclusion of DAT terms in the discrete lattice Hamiltonian, the model is now generally termed as {\em non-standard} BHM \cite{dut15}.


\subsection{Rydberg admixed potential: density-density interactions}
\label{bryd}
\begin{figure}[hbt]
\centering
\includegraphics[width= .9\columnwidth]{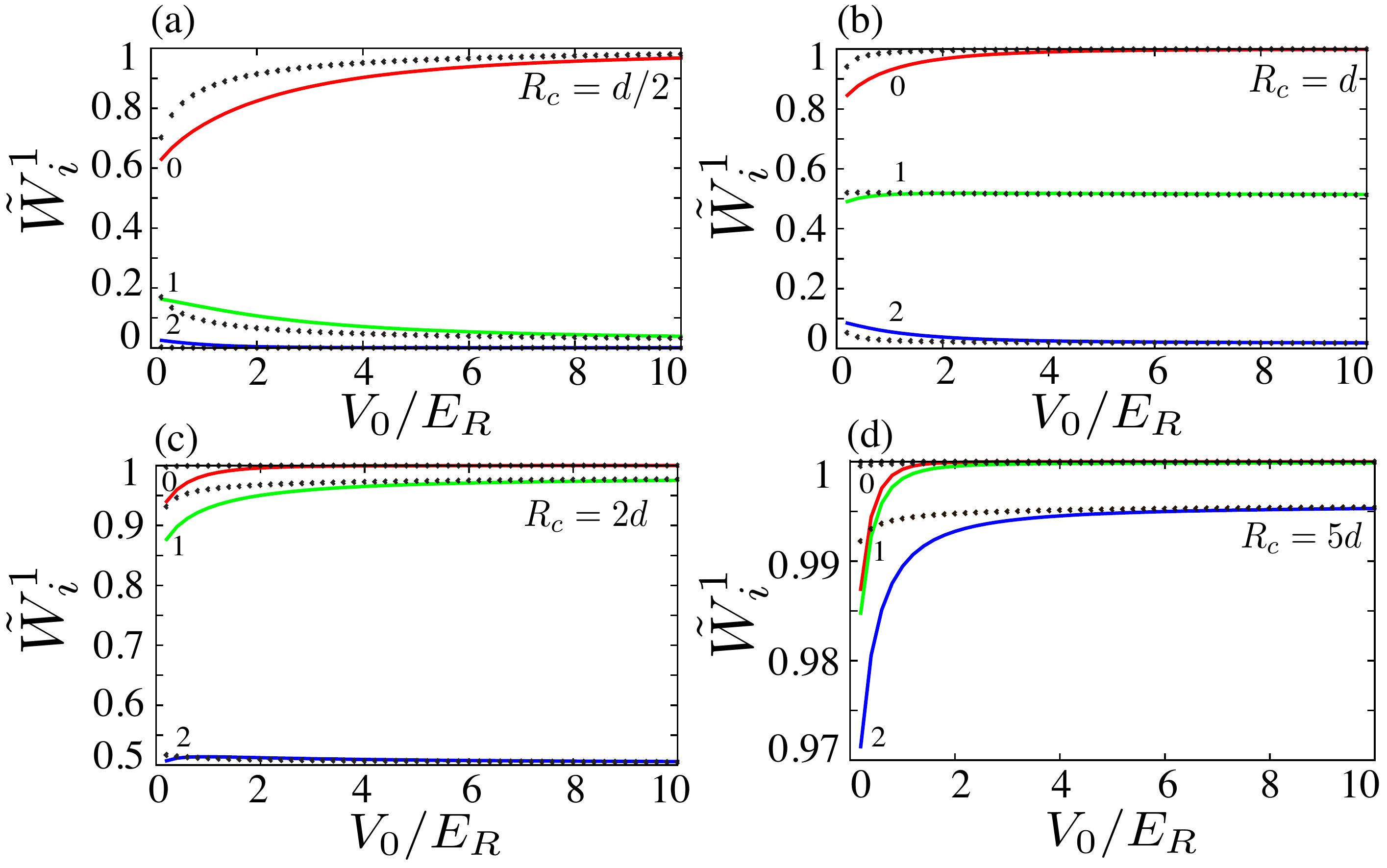}
\caption{\small{The Rydberg admixed interaction parameters $\tilde W_i^1=W_{0i0i}^{1111}/W_{eff}$ in the lowest band ($n=m=1$) as a function of the lattice depth $V_0$ with (a) $R_c=d/2$, (b) $R_c=d$, (c) $R_c=2d$ and (d) $R_c=5d$. The onsite ($i=0$), nearest neighbour ($i=1$) and next nearest neighbour ($i=2$) interaction couplings are shown, indicated by the numbers in the plot. As $R_c$ increases, the off-site matrix elements become significant as that of onsite ones. This arises from the flattened nature of the soft-core barrier of the Rydberg induced potential. In addition, they all saturate to the same value independent of the separation between the atoms as $R_c$ becomes larger and larger. The results from the tight binding approximation are shown in dotted lines. }}
\label{fig:b1} 
\end{figure}
We rewrite the binary interaction (equation \ref{dpot}) between the Rydberg admixed atoms as 
\begin{equation}
W(r_{ij})=\frac{W_{eff}}{\left[(r_{ij}/R_c)^6+1\right]},
\label{dpotn}
\end{equation}
 with $W_{eff}=\hbar\Omega^4/(8|\Delta|^3)$ being an effective interaction strength given by the light shift due to the Rydberg laser and $R_c=[C_6/2\hbar|\Delta|]^{1/6}$ provides the interaction range. Among the two parameters, $R_c$ can be controlled independently of $W_{eff}$ by changing the Rydberg state through $C_6(n)$, but any variation in $\Delta$ and $\Omega$ affects both simultaneously. The possibility of tuning the interaction range without affecting the effective strength can be pointed out as an  elite feature of Rydberg admixed atoms compared to other existing long-range systems. See section. \ref{rb} for the values of $R_c$ as a function of $C_6(n)$ (or simply $n$) for Rubidium atoms. 
 
 \begin{figure}[hbt]
\centering
\includegraphics[width= .9\columnwidth]{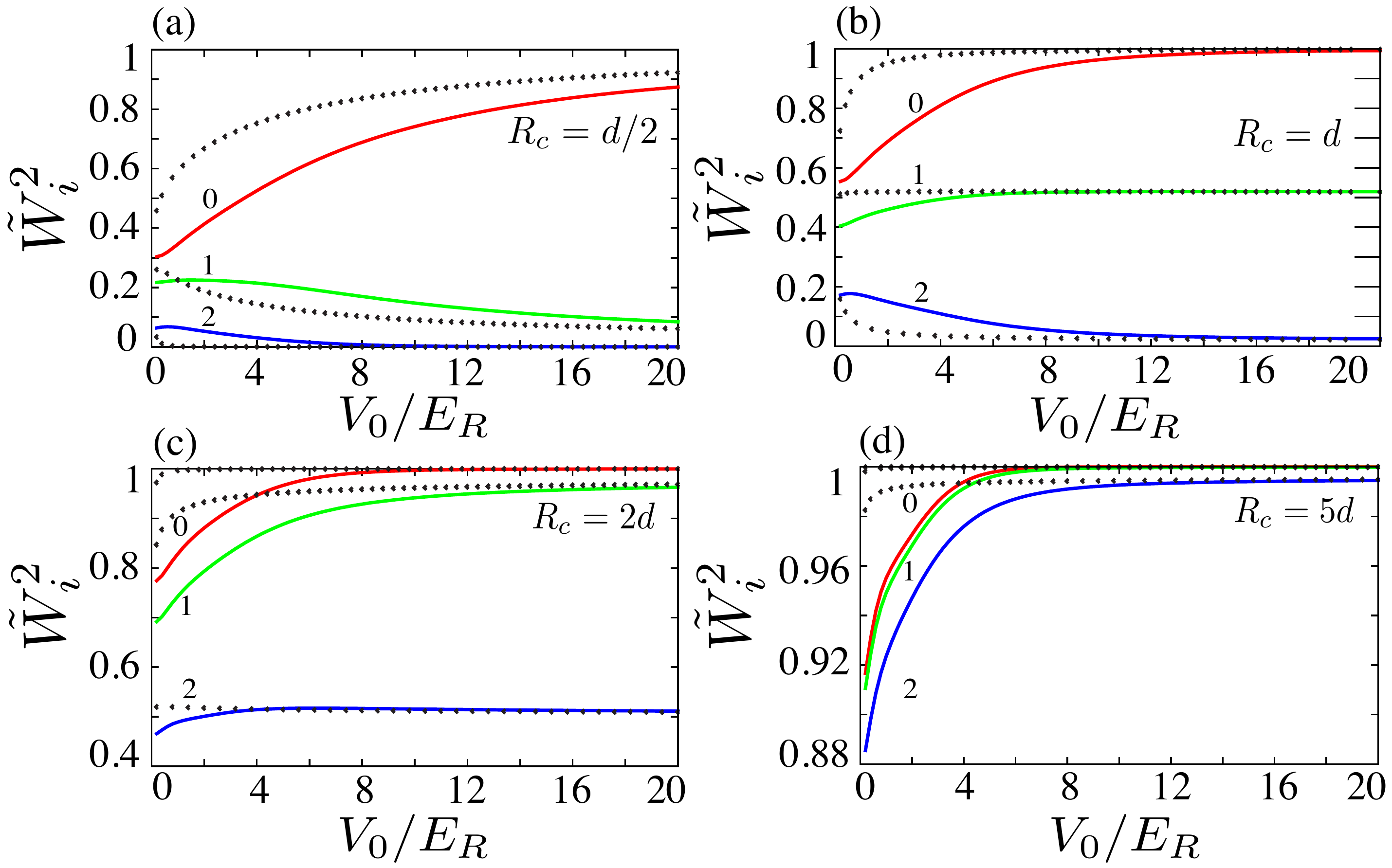}
\caption{\small{The same as in figure \ref{fig:b1}, but for the first excited band ($\tilde W_i^2=W_{0i0i}^{2222}/W_{eff}$). The results from tight binding approximation are shown by the dotted lines. The value of $i$ is used in labelling the plots.}}
\label{fig:b2} 
\end{figure}

\begin{figure}[hbt]
\centering
\includegraphics[width= .9\columnwidth]{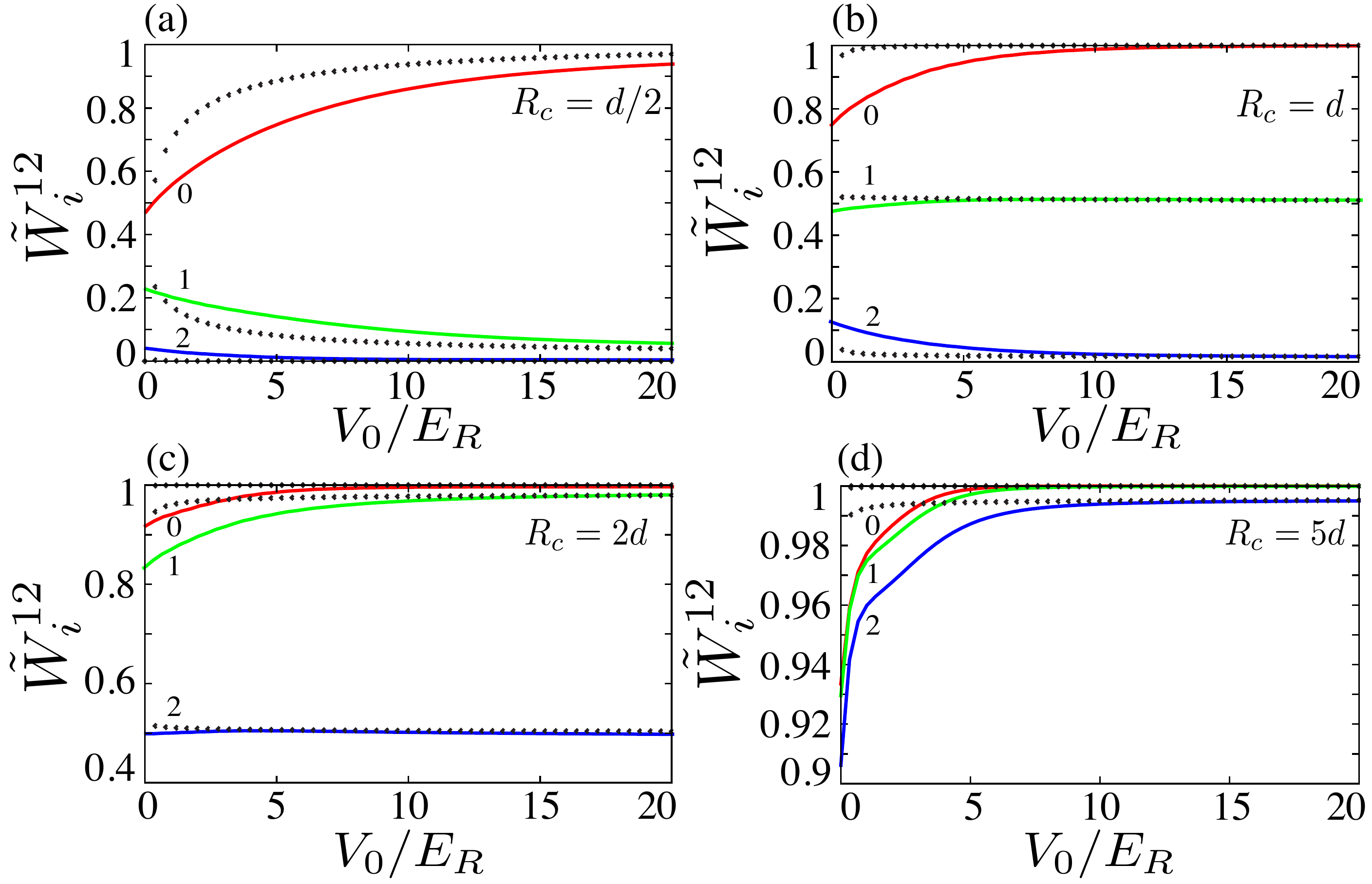}
\caption{\small{The inter-band  matrix elements ($\tilde W_i^{12}=W_{0i0i}^{1212}/W_{eff}$) for the Rydberg admixed interaction as a function of lattice depth $V_0$ for (a) $R_c=d/2$, (b) $R_c=d$, (c) $R_c=2d$ and (d) $R_c=5d$. The results from tight binding approximation are shown by the dotted lines. The value of $i$ is used in labelling the plots.}}
\label{fig:b12} 
\end{figure}

 With this, we estimate the Hubbard parameters for Rydberg-dressed interaction ($W_{iji'j'}^{nmn'm'}$) using MLWFs, as a function of both lattice depth $V_0$ and $R_c$. The results for the case in which two atoms occupy the lowest band ($\tilde W_i^1=W_{0i0i}^{1111}/W_{eff}$) are shown in figure \ref{fig:b1}. For any value of $R_c$, the onsite interaction ($i=0$) increases with increasing $V_0$ but it saturates rather quickly compared to that of short-range interactions for sufficiently large values of $R_c$. This is attributed to the flat nature of the softcore potential. The behaviour of off-site matrix elements with $V_0$ depends crucially on the value of $R_c$. For instance, for $R_c=d/2$ all of them decreases with $V_0$ and become negligible for sufficiently large $V_0$ [figure \ref{fig:b1}(a)]. When $R_c\sim d$, the character of the nearest neighbour ($i=1$) coupling changes and it increases with $V_0$ [figure \ref{fig:b1}(b)]. For $R_c\sim 2d$ it becomes almost identical to that of the onsite interaction [figure \ref{fig:b1}(c)] and in addition, the next nearest neighbour coupling becomes significantly large, becomes half of onsite and first nearest neighbour interactions. As shown in figure \ref{fig:b1}(d) for $R_c=5d$ all the three are almost same in magnitude. Hence, as $R_c$ increases the off-site matrix elements become as relevant as onsite interactions. The same results as above for the first excited band are shown in the figure. \ref{fig:b2} and that of when one atom occupying the lowest band and the second in the first excited band is shown in the figure. \ref{fig:b12}. For all cases, the tight binding results are shown in dotted lines and they agree very well with the exact calculations when $V_0$ is sufficiently large. For $V_0\gg E_R$, the Wannier states can be approximated to the Dirac delta functions and in this limit the interactions ($W_i^1$ and $W_i^2$) are just given by the bare potential $W(x)$.

 The important properties to be noted as follows: the first point is that due to the flattened nature of the Rydberg potential at short distances, the dominant matrix elements for the density-density interactions saturate quickly to $W_{eff}$ as a function of $V_0$ when $R_c>d$, . This can be seen again as a unique feature of Rydberg-dressed interactions compared to other inter-atomic potentials. This is because the potential $W(x-x^{\prime})$ changes hardly over an inter-particle distance of $x-x^{\prime}\sim R_c$, hence it can be approximated to a constant such that  
 \begin{equation}
W_{0i0i}^{nmnm}\simeq W(x-x')\int dx  \ |w_n^*(x)|^2 \int dx' \  |w_m^*(x'-X_i)|^2=W_{eff}.
\label{int1}
\end{equation}
As we go higher in the bands, the saturation behaviour with $V_0$ slows down. It can also be seen as that the saturation occurs when the MLWFs are fully accommodated inside the soft-core barrier of the potential or when $\Omega < R_c$ and it delays the saturation as a function of $V_0$ for higher bands. The second point is that nearest neighbouring interactions can be made same strength as that of onsite interactions by adjusting the value of $R_c$ i.e. they become position-independent. This may have far-reaching consequences in the context of many body physics \cite{matt13}, and in particular on frustrated magnetism \cite{ale14, ale15}, to impose local constraints. Since $R_c$ is varied through vdW coefficient $C_6$, the different many body quantum phases arising from the long-range nature of the interactions can be accessed simply as a function of the principal quantum number $n$ without changing other parameters in the system. The quantitative details for a Rubidium lattice setup are worked out in section. \ref{rb}.

 \begin{figure}[hbt]
\centering
\includegraphics[width= .7\columnwidth]{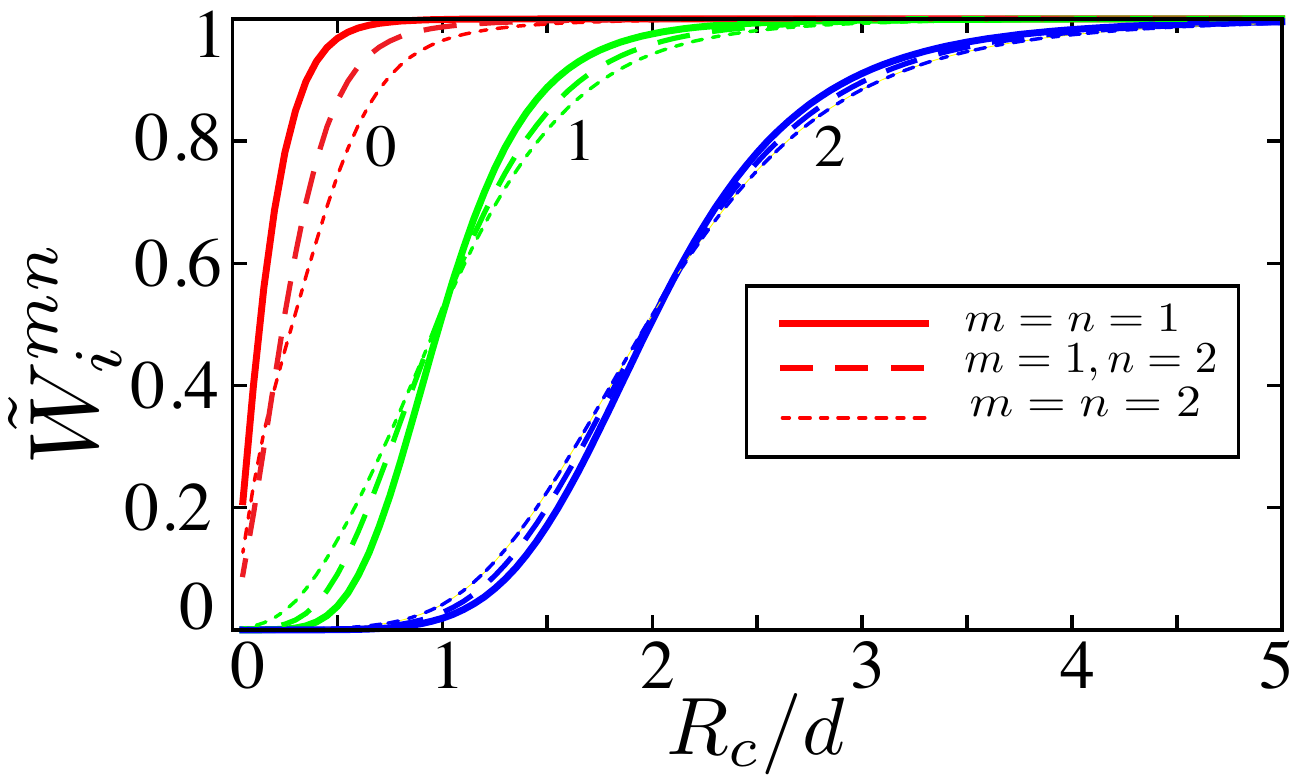}
\caption{\small{The onsite and off-site matrix elements ($\tilde W_i^{mn}=W_{0i0i}^{mnmn}/W_{eff}$) for the Rydberg admixed interactions as a function of $R_c$ for $V_0=10 E_R$. The numbering shown in the plots are for onsite ($i=0$) , nearest neighbour ($i=1$) and next nearest neighbours ($i=2$). For each of them the results are shown for the lowest band $(m=n=1)$, first excited band ($m=n=2$) and between them $(m=1, n=2)$. For $i=1$ and $i=2$ there is a crossing between the parameters for the lowest and the first excited bands (see text).}}
\label{fig:urc} 
\end{figure}

 We summarize this section by showing the interaction parameters up to second nearest neighbours in the first two bands as a function of $R_c$ for a fixed $V_0$ [figure \ref{fig:urc}]. The saturation of matrix elements are clearly visible here as well, and can be explained the same way as we did before. Another feature we have seen as that the off-site matrix elements for the lowest and the first excited bands cross each other. At small $R_c$, it is the overlap between the MLWFs which determines the off-site couplings and they are larger for first excited band compared to the lowest one. On the contrary, at sufficiently large $R_c$ the off-site coupling strength depends on how well it is accommodated inside the soft-core barrier and in that case the lowest band dominates resulting in a crossing between them as a function of $R_c$. We also noticed that  these crossings are even more prominent at small values of $V_0$.

\subsection{Rydberg admixed potential: density assisted tunnelings}

\begin{figure}[hbt]
\centering
\includegraphics[width= .9\columnwidth]{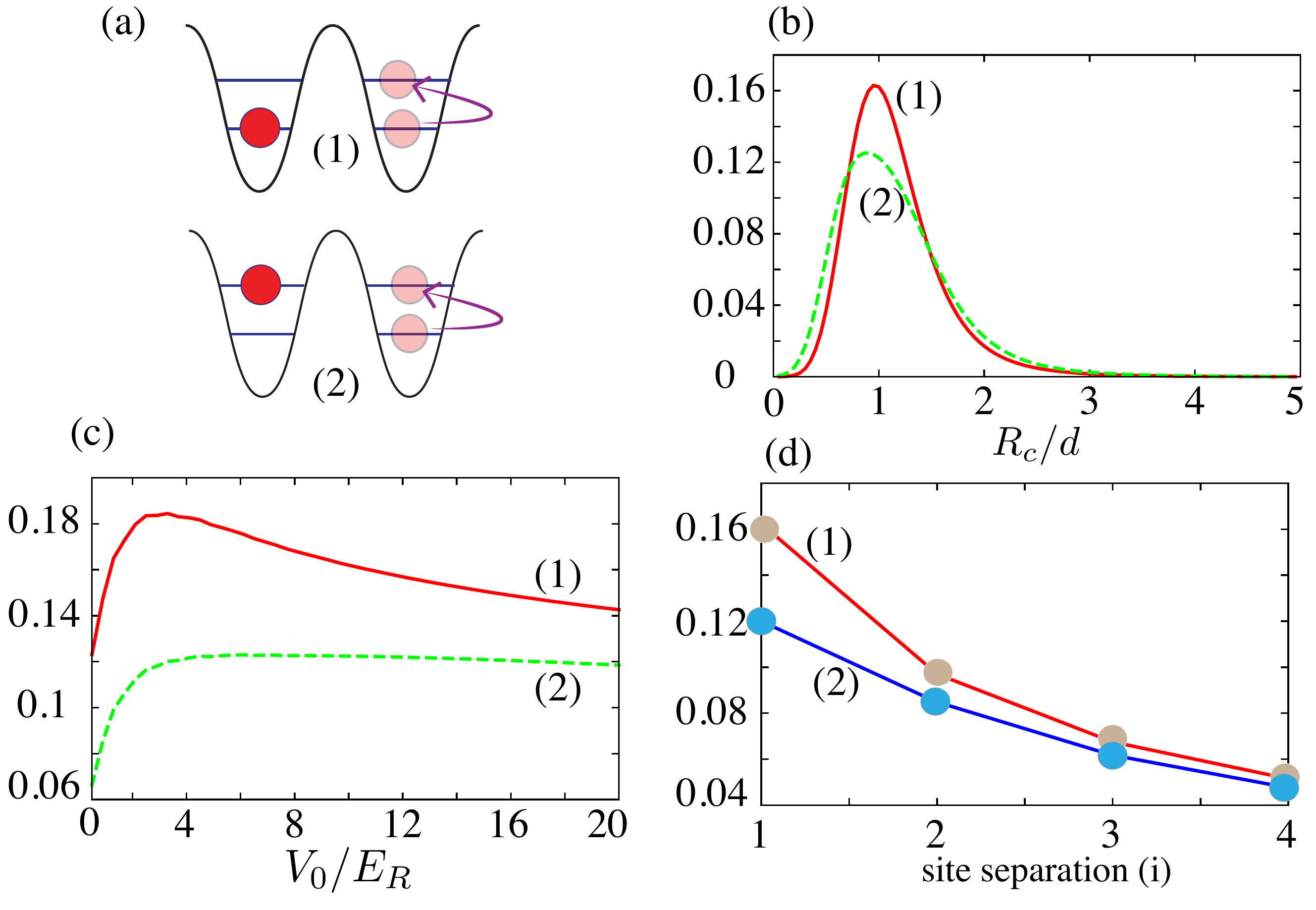}
\caption{\small{(a) The two dominant off-site DAT inter-band processes due to the Rydberg admixed interactions. (b) The amplitudes or strengths of the two events as a function of $R_c$ for $V_0=10 E_R$. Note that for the nearest neighbour it is peaked at $R_c=d$, and if next nearest neighbours are considered the peak will appear around $R_c=2d$ but with a lesser magnitude. (c) The same as a function of $V_0$ for $R_c=d$. (d) The maximum of the peak appearing in (c) as a function of inter-site separation for $V_0=10 E_R$. $i=1$ means nearest neighbour and so on. Note that the peak gets shifted to $i\times R_c$ as a function of $i$. The scaled amplitudes (units of $W_{eff}$) are plotted along the y-axis for figures (b)-(d) and they are labelled by the numbers indicated for the two processes.}}
\label{fig:dat2} 
\end{figure}

We look at the DAT processes induced by the Rydberg admixed interactions. Here, we only discuss the two dominant processes [schematically depicted in figure \ref{fig:dat2} (a)] with in the two lowest Bloch bands and they both are found to be DAT inter-band tunnelings, with an explicit dependence on $R_c$. For a given $R_c$ and $V_0$, the amplitudes of these processes as a function of inter-particle separation exhibits a sharp peak at $R_c$. That means, if $R_c=d$ only for the nearest neighbour atoms these processes takesplace, and similarly when $R_c=2d$ it happens between the next nearest neighbour atoms.

 Let us now focus on the case of nearest neighbours [depicted in figure \ref{fig:dat2} (a)]. For large $R_c$, the Rydberg-admixed potential can be taken as a constant and it results in: $\sim\int w_0(x)w_1^*(x) dx \approx 0$, hence, you observe a rapid decay for $R_c>d$. For $R_c<d$, the interactions between the sites are weak enough to trigger this processes and hence resulting in a hump like behaviour as a function of $R_c$.  The same matrix elements as a function of $V_0$ for a given $R_c(=d)$ are shown in figure \ref{fig:dat2} (c).  If we consider the same processes between the next nearest neighbours, the above-mentioned peak appears at $R_c=2d$ but with a lesser magnitude and so on. Hence, we calculated the maximum of the peak as a function of inter-particle separation ($R_c$ has changed accordingly) for $V_0=10$ $E_R$ [see figure \ref{fig:dat2} (d)]. Note that the amplitudes for all other DAT events are at least one order of magnitude lower than those shown in  figure \ref{fig:dat2}(a). The above processes introduces the following type of terms: $\sim \sum_{i,j}\hat a^{\dagger}_{j2}\hat n_{i1}\hat a_{j1}$  [process 1 in figure \ref{fig:dat2}(a)] and $\sim \sum_{i,j}\hat a^{\dagger}_{j2}\hat n_{i2}\hat a_{j1}$ [process 2 in figure \ref{fig:dat2}(a)] with in the two lowest-band EBHM for the Rydberg-admixed atoms in a 1D optical lattice.

\subsection{An example: Rubidium atoms}
\label{rb}
\begin{figure}[hbt]
\centering
\includegraphics[width= .9\columnwidth]{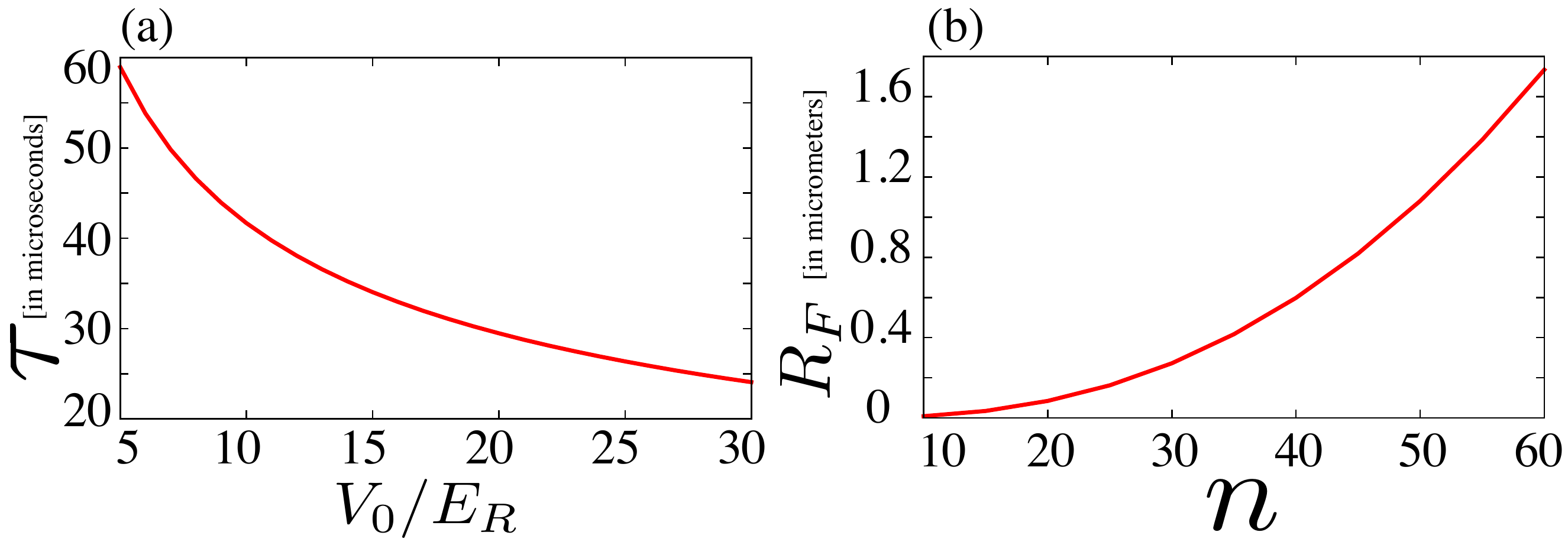}
\caption{\small{(a) The time ($\tau$) taken for the atomic wave packet to increment $10\%$ of its initial width during free expansion as a function of $V_0$. (b) Numerical calculations for the critical distance at which the F\"orster and vdW regimes are separated as a function of $n$ for Rubidium $nS_{1/2}$ states.}}
\label{fig:crit} 
\end{figure}

In this section we consider the state of the art Rubidium ($^{87}Rb$) lattice setup. If the Rydberg state atoms are unconfined, as discussed in Section \ref{val} the initial atomic (Wannier) wave packet for the ground state atoms undergoes free expansion resulting in the de-localization of the particle. This disturbs the picture of the localized Hubbard model. In order to preserve that, the experimental process has to be restricted with in the time scale at which the spatial extension of the initial wave packet remains unchanged. The later is determined by the Equation \ref{cri} and in figure \ref{fig:crit}(a) we plot the time taken in which the width of the wave packet is incremented by $10\%$ from the initial value as a function of $V_0$, for Rubidium atoms with a lattice spacing of $d=4 \ \mu m$. The lattice spacing of $4\ \mu m$ can be achieved with a co-propagating laser beams, of wavelength $\lambda=1064$ nm, forming an  angle of $\theta=13.3$ degrees between them. As $V_0$ increases the initial width gets lesser and lesser and the wave packet carries more kinetic energy hence, takes lesser time to expand. This definitely restricts the study of Hubbard-like model with un-trapped Rydberg states with large principal quantum numbers in deep optical lattices. For $V_0=10E_R$, the maximum value of $n$ one can attain is $\sim 35$. Therefore magic wavelength optical lattices are strongly recommended to explore the physics in the tight-binding regimes using Rydberg-dressed atoms involving large principal quantum numbers.

 On the other hand for any given Rydberg state $nS_{1/2}$, the validity of the vdW picture in our lattice model requires that the lattice separation must be larger than $R_F$ [see Equation (\ref{rc})], the cross-over point between the F\"orster regime and vdW regime. In figure \ref{fig:crit}(b), we show the numerical results of $R_F$ as a function of $n$ for $nS_{1/2}$ Rubidium states. Though the main contributions from the dipole coupling are coming from the nearest $nP_{1/2}$, we also included $nP_{3/2}$. It shows that in order to acsess the vdW regime the lattice spacings are at least of about $2\mu m$ for $n>60$.

\label{rb}
\begin{figure}[hbt]
\centering
\includegraphics[width= .9\columnwidth]{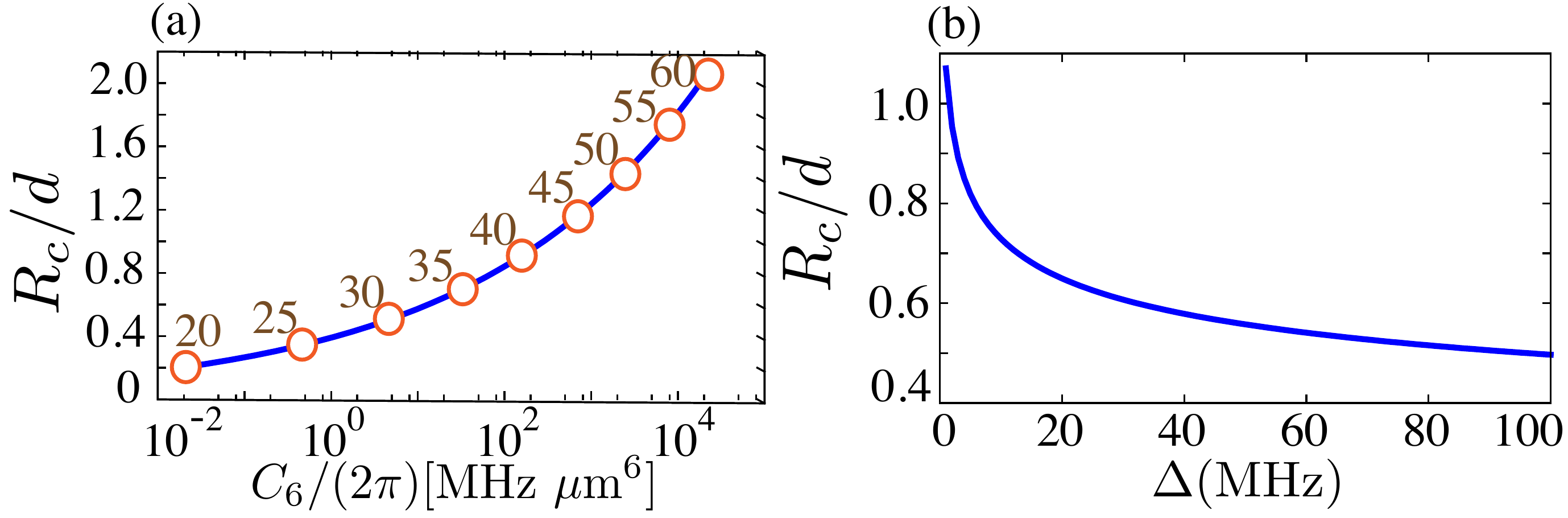}
\caption{\small{(a) The range of Rydberg admixed interaction ($R_c$) as a function of $C_6(n)$ for $^{87}Rb$ atoms in which the ground state is weakly admixed to $nS_{1/2}$ states with a detuning $\Delta= 100$ MHz and a lattice space of $d=2\mu m$. The corresponding principal quantum numbers $n$ are indicated by the empty bubbles. (b) $R_c$ as a function of $\Delta$ for $30S_{1/2}$.}}
\label{fig:rc} 
\end{figure}

As we have pointed out earlier in section \ref{bryd}, any change in detuning $\Delta$ or $\Omega$ affect $R_c$ and $W_{eff}$ simultaneously, but the interaction range $R_c$ can be varied independently by changing the Rydberg states. Below we estimate realistic values for $R_c$ as a function of $C_6(n)$ for Rubidium  atoms  weakly coupled to $nS_{1/2}$ states, in an 1D optical lattice [figure \ref{fig:rc}(a)]. We consider a lattice spacing of $d=2 \mu m$. It can be seen that the range of interactions can be extended over as many as 2 sites when considering a Rydberg state $60S_{1/2}$ with $\Delta=100$ MHz. Note that for fixed $C_6$ the variation of $R_c$ against $\Delta$ is relatively slow. With $\Omega/\Delta=0.1$ and $\Delta= 100$ MHz, we can attain an effective interaction strength of $W_{eff}=10$ kHz.

\section{Conclusion}
\label{con}
In conclusion, we have estimated {\em ab initio} the Hubbard parameters for Rydberg-dressed atoms in an one-dimensional sinusoidal optical lattice in the basis of maximally localized Wannier states. The finite range, soft-core nature of the Rydberg induced vdW interactions strongly modifies the nature of Hubbard parameters. In particular, in controlling the spatial extension of the soft-core barrier one can engineer the off-site interactions and make them same magnitude as that of the onsite ones, hence making them position independent. This may influence the many body dynamics in a great deal. In addition to the typical density density interactions, we provide the information for the DAT coefficients in this system. Two dominant process which depends crucially on the soft-core have been identified and discussed which will be highly relevant when studying non-standard multi band Hubbard models. In the end, we have discussed realistic parameters for the state of the art experimental Rydberg dressed Rubidium lattice setup.
\section{Acknowledgements}
We acknowledge useful discussions with Venkateswara Pai, Prasenjit Ghosh, Weibin Li and  Umesh Waghmare. R.N acknowledges the funding by the Indo-French Centre for the Promotion of Advanced Research (CEFIPRA).

\appendix
\addcontentsline{toc}{section}{Appendices}
\addtocontents{toc}{\protect\setcounter{tocdepth}{-1}}

\section{Single particle energy bands and Bloch states in an 1D optical lattice}
\label{bbs}
The Hamiltonian for a single particle in the presence of an 1D optical lattice potential $V(x)$ is
\begin{equation}
\hat h_0=-\frac{\hbar^2}{2m}\frac{d^2}{dx^2}+V_{OL}(x),
\end{equation}
where $V_{OL}(x)=V_0\sin^2(k_0x)$ and the wave number $k_0=\pi/d$ with $d$ being the lattice spacing.
The single particle energy band structure is obtained by solving the eigen-value equation $\hat h_0\psi_q^n(x)=E_q^n\psi_q^n(x)$ for a given quasi-momentum $q\in [-\pi/d,\pi/d]$ and $n$ is the band index.  According to Bloch theorem, we have Bloch solutions $\psi_q^n(x)=e^{iqx}u_n^q(x)$ and $u_n^q(x)$ has the same periodicity as that of the lattice potential. The above eigen value equation is best tackled in the Fourier space, and by introducing the Fourier expansion for the lattice potential $V_{OL}(x)=(1/\sqrt{d})\sum_kv_ke^{iG_kx}$ and for $u_q^n(x)=(1/\sqrt{d})\sum_kc_{q,k}^ne^{iG_kx}$. With Fourier coefficients $c_{q,k}^n$ satisfying the normalization condition $\sum_k(c_{q,k}^m)^*c_{q,k}^n=\delta_{m,n}$, the periodic functions $u_q^n(x)$ is normalized over the primitive unit cell of the lattice in the real space. Finally we arrive at

\begin{equation}
\frac{\hbar^2}{2m}\left(G_k+q\right)^2c_{q,k}^n+\frac{1}{\sqrt{d}}\sum_{k^{\prime}}v_{k-k^{\prime}}c_{q,k^{\prime}}^n=E_q^nc_{q,k}^n.
\end{equation}
We can write the above equation in a matrix form as $H_qc_q^n=E_q^nc_q^n$ with eigen vectors $c_q^n$ and eigen values $E_q^n$. The matrix elements of $H_q$:
\begin{equation}
(H_q)_{ij}=\frac{\hbar^2}{2m}\left(G_i+q\right)^2\delta_{ij}+\frac{1}{\sqrt{d}}v_{i-j}.
\label{hmat}
\end{equation}
Diagonalizing $H_q$ provides us the Fourier coefficients $c_{q,k}^n$ and they possess certain properties based on various symmetries. If inversion symmetry exists and together with time-reversal symmetry, they guarantee that $c_{q,k}^n$ are real apart from a common phase factor and also satisfies $c_{q,k}^n=(c_{-q,-k}^n)^*$. This property can considerably reduce the numerical cost in calculating the Bloch functions $\psi_q^n(x)$. 
We employ a LAPACK routine to obtain the eigen states and eigen values of the Hamiltonian matrix given in equation \ref{hmat} for $q\in [-\pi/d,\pi/d]$. The eigen vectors  $c^n_{q}$ are real-valued and the periodic Bloch function $u_q^n(x)$ is then obtained by its Fourier transform.
\section*{References}

\providecommand{\newblock}{}

\begin{thebibliography}{10}
\expandafter\ifx\csname url\endcsname\relax
  \def\url#1{{\tt #1}}\fi
\expandafter\ifx\csname urlprefix\endcsname\relax\def\urlprefix{URL }\fi
\providecommand{\eprint}[2][]{\url{#2}}

\bibitem{sant00} Santos L, Shlyapnikov G V, Zoller P, and Lewenstein M 2000 {\em Phys. Rev. Lett.} {\bf 85} 1791
\bibitem{hon10} Honer J, Weimer H, Pfau T and B\"uchler H P, 2010 {\em Phys. Rev. Lett.} {\bf 105} 160404
\bibitem{john10} Johnson J E and Rolston S L, 2010 {\em Phys. Rev. A} {\bf 82} 033412
\bibitem{pup10} Puppilo G, Micheli A, Boninsegni, Lesanovsky I and Zoller P 2010 {\em Phys. Rev. Lett.} {\bf 104} 223002
\bibitem{hen10} Henkel N, Nath R and Pohl T 2010 {\em Phys. Rev. Lett.} {\bf 104} 195302
\bibitem{tom14} Macr\`i T and Pohl T 2014  {\em Phys. Rev. A} {\bf 89} 011402
\bibitem{bal14} Balewski J B, Krupp A T, Gaj A, Hofferberth S, L\"ow R and Pfau T 2014 {\em New J. Phys.} {\bf 16} 063012
\bibitem{ric15} Mukherjee R, Killian T C and Hazzard R A, arXiv:1511.08856
\bibitem{saf10} Saffman M, Walker T G and Molmer K 2010 {\em Rev. Mod. Phys.} {\bf 82} 2313
\bibitem{luk01} Lukin M D, Fleischhauer, Cote R, Duan L M, Jaksch D, Cirac J I and Zoller P 2001 {\em Phys. Rev. Lett.} {\bf 87} 037901
\bibitem{urb09} Urban E, Johnson T A, Henage T, Isenhower L, Yavuz D D, Walker T G and Saffman M 2009 {\em Nature Phys.} {\bf 5} 110
\bibitem{gae09} Ga\"etan A, Miroshnychenko Y, Wilk T, Chotia A, Viteau M, Comparat D, Pillet P, Browaeys A and Grangier P 2009 {\em Nature Phys.} {\bf 5} 115
\bibitem{poh10} Pohl T, Demler E and Lukin M D, 2010 {\em Phys. Rev. Lett.} {\bf 104} 043002
\bibitem{wei08} Weimer H, L\"ow R, Pfau T and B\"uchler H P 2008  {\em Phys. Rev. Lett.} {\bf 101} 250601
\bibitem{low09} L\"ow R, Weimer H, Krohn U, Heidemann R, Bendkowsky V, Butscher B, B\"uchler H P, and Pfau T 2009 {\em Phys. Rev. A} {\bf 80} 033422
\bibitem{wei10}Weimer H and B\"uchler H P, 2010 {\em Phys. Rev. Lett.} {\bf 105} 230403
\bibitem{olm09}Olmos B, Gonz\'alez-F\'erez, and I. Lesanovsky 2009 {\em  Phys. Rev. Lett.}{\bf 103} 185302
\bibitem{ji11} Ji S, Ates C and Lesanovsky I 2011 {\em  Phys. Rev. Lett.} {\bf 107} 060406
\bibitem{igo11} Lesanovsky I 2011 {\em Phys. Rev. Lett.} {\bf 106} 025301
\bibitem{jau15} Jau Y Y, Hankin A M, Keating T, Deutsch I H and Biedermann G W 2015, {\em Nature Phys. }
\bibitem{zei16} Zeiher J, van Bijnen R M W, Schauss P, Hild S, Choi J-Y, Pohl T, Bloch I and Gross C, {\em arXiv:1602.06313}
\bibitem{hsu12}Hsueh C H, Lin T C, Horng T L and Wu W C 2012 {\em Phys. Rev. A}, {\bf 86} 013619
\bibitem{cin10} Cinti F, Jain P, Boninsegni M, Micheli A, Zoller P and Pupillo G 2010 {\em Phys. Rev. Lett.} {\bf 105} 135301
\bibitem{li15} Li S and Sarma S D 2015 {\em Nature Comm.} {\bf 6} 7137
\bibitem{gei15} Geissler A, Vasic I and Hofstetter W, arXiv:1509.06292
\bibitem{mau11} Maucher F, Henkel N, Saffman M, Kr\'olikowski W, Skupin S and Pohl T 2011 {\em Phys. Rev. Lett.} {\bf 106} 170401
\bibitem{ric15} van Bijnen B M W and Pohl T 2015 {\em Phys. Rev. Lett.} {\bf 114} 243002
\bibitem{dau12} Dauphin A, M\"uller M, and Martin-Delgado M A 2012, {\em Phys. Rev. A} {\bf 86} 053618
\bibitem{ale14} Glaetzle A W, Dalmonte M, Nath R, Rousochatzakis I, Moessner R and Zoller P 2014 {\em Phys. Rev. X} {\bf 4} 041307
\bibitem{ale15} Glaetzle A W, Dalmonte M, Nath R, Gross C, Bloch I and Zoller P 2015 {\em Phys. Rev. Lett.} {\bf 114} 173002
\bibitem{kea13}Keating T, Goyal K, Jau Y Y, Biedermann G W, Landahl A J and Deutsch I H 2013 {\em Phys. Rev. A} {\bf 87} 052314
\bibitem{kea15}Keating T, Cook R L, Hankin A M, Jau Y Y, Biedermann G W, and Deutsch I H 2015 {\em Phys. Rev. A} {\bf 91} 012337
\bibitem{mob13}M\"obius S, Genkin M, Eisfeld A, Wüster S, and Rost J M 2013 {\em Phys. Rev. A} {\bf 87} 051602(R)
\bibitem{gil14}Gil L I R, Mukherjee R, Bridge E M, Jones M P A and Pohl T 2014 {\em Phys. Rev. Lett.}, {\bf 112}, 103601
\bibitem{bou02} Bouchoule I and Mølmer K 2002, {\em Phys. Rev. A}, {\bf 65}, 041803
\bibitem{gau15} Gaul C, DeSalvo B J, Aman J A, Dunning F B, Killian T C and Pohl T arXiv:1511.06424
\bibitem{lew12} Lewenstein M, Sanpera A and Ahufinger V 2012 {\em Ultracold atoms in optical lattices: Simulating quantum many body systems} (Oxford University press, New York, 2012).
\bibitem{blo08}Bloch I, Dalibard J, and Zwerger W 2008 {\em Rev. Mod. Phys.} {\bf 80} 885
\bibitem{jak98}Jaksch D, Bruder D, Cirac J I, Gardiner C W and Zoller P 1998 {\em Phys. Rev. Lett.}, {\bf 81}, 3108
\bibitem{fis89}Fisher M P A, Weichman P B, Grinstein G, and Fisher D S 1989 {\em Phys. Rev. B}, {\bf 40} 546
\bibitem{gre02}Greiner M, Mandel O, Esslinger T, H\"ansch T W and Bloch I 2002 {Nature}, {\bf 415} 39
\bibitem{ger08}Gericke T, Würtz P, Reitz D, Langen T  and Ott H 2008 {\em Nature Phys.}, {\bf 4}, 949 
\bibitem{bak09}Bakr W S, Gillen J I, Peng A, Fölling S and Greiner M 2009 {\em Nature}, {\bf 462}, 74
\bibitem{she10}Sherson J F, Weitenberg C, Endres M, Cheneau M, Bloch I and Kuhr S 2010 {\em Nature}, {\bf 467} 68
\bibitem{koh59}Kohn W 1959 {\em Phys. Rev.},{\bf 115} 809
\bibitem{bro07}Brouder C, Panati G, Calandra M, Mourougane C, and Marzari N 2007 {\em Phys. Rev. Lett.}, {\bf 98}, 046402
\bibitem{mar97}Marzari N and Vanderbilt D 1997, {\em Phys. Rev. B}, {\bf 56}, 12847
\bibitem{mar12}Marzari N, Mostofi A A, Yates J R, Souza I and Vanderbilt D 2012 {\em Rev. Mod. Phys.}, {\bf 84} 1419
\bibitem{wal13}Walters R, Cotugno G, Johnson T H, Clark S R, and Jaksch D 2013 {\em Phys. Rev. A}, {\bf 87}, 043613
\bibitem{bar12}Baranov M, Dalmonte M, Pupillo G, and Zoller P 2012, {\em Chem. Rev.} {\bf 112}, 5012 
\bibitem{tre11}Trefzger C, Menotti C, Capogrosso-Sansone B and Lewenstein M 2011, {\em J. Phys. B: At. Mol. Opt. Phys.}, {\bf 44}, 193001
\bibitem{sca95} Scalettar R T, Batrouni G G, Kampf A P, and Zimanyi G T 1995, {Phys. Rev. B}, {\bf 51} 8467
\bibitem{sca05}Scarola V W and Sarma S D 2005, {\em Phys. Rev. Lett.}, {\bf 95}, 033003
\bibitem{yi07} Yi S, Li T, and Sun C P 2007, {\em Phys. Rev. Lett.}, {\bf 98}, 260405
\bibitem{pol10}Pollet L, Picon J D, Büchler H P, and Troyer M 2010 {\em Phys. Rev. Lett.}, {\bf 104}, 125301
\bibitem{cap10}Capogrosso-Sansone B, Trefzger C, Lewenstein M, Zoller P, and Pupillo G 2010, {\em Phys. Rev. Lett.}, {\bf 104} 125301
\bibitem{dall06} Dalla Torre E G, Berg E, and Altman E 2006, {\em Phys. Rev. Lett.}, {\bf 97} 260401
\bibitem{bai15}Baier S, Mark M J, Petter D, Aikawa K, Chomaz L, Cai Z, Baranov M, Zoller P and Ferlaino F arXiv:1507.03500
\bibitem{jur14}J\"urgensen, Meinert F, Mark M J, N\"agerl H C, and L\"uhmann D S 2014 {\em Phys. Rev. Lett.} {\bf 113} 193003
\bibitem{matt13}Mattioli M, Dalmonte M, Lechner W, and Pupillo G 2013 {\em Phys. Rev. Lett.}, {\bf 111} 165302
\bibitem{dut15} Dutta O, Gajda M, Hauke P, Lewenstein M, L\"uhmann D.-S, Malomed B A, Sowi\'nski T and Zakrzewski J 2015 {\em Rep. Prog. Phys.} {\bf 78} 066001
\bibitem{sow12}Sowi\'nski T, Dutta O, Hauke P, Tagliacozzo T, and Lewenstein M 2012 {\em Phys. Rev. Lett.} {\bf 108} 115301
\bibitem{saf03} Safronova M S, William C J and Clark C W 2003 {\em Phys. Rev. A} {\bf 67} 040303
\bibitem{saf05}Saffman S, and Wlaker T G 2005 {\em Phys. Rev. A} {\bf 72} 022347
\bibitem{muk11}Mukherjee R, Millen J, Nath R, Jones M P A and Pohl T 2011 {\em J. Phys. B: At. Mol. Opt. Phys.} {\bf 44} 184010
\bibitem{gol15}Goldschmidt E A, Norris D G, Koller S B, Wyllie R, Brown R C, Porto J V, Safronova U I and Safronova M S 2015 {\em Phys. Rev. A} {\bf 91} 032518
\bibitem{top13}Topcu T and Derevianko 2013 {\em Phys. Rev. A} {\bf 88} 043407
\bibitem{li13} Li W, Ates C and Lesanovsky I 2013 {\em Phys. Rev. Lett.} {\bf 110} 213005
\bibitem{nel07} Nelson D K, Li X and Weiss D S 2007 {\em Nat. Phys.}{\bf 3} 556
\bibitem{he01}He L and Vanderbilt D 2001 {\em Phys. Rev. Lett.}, {\bf 86} 5341
\bibitem{edm63} Edmiston C and Ruedenberg K 1963 {\em Rev. Mod. Phys.}, {\bf 35} 457
\bibitem{wal12} Walters R 2012 {\em Ultra-cold atoms in optical lattices: Simulating quantum spin systems} PhD thesis.
 \end{thebibliography}

\end{document}